\documentclass[10pt,preprint]{aastex}

\slugcomment{To Appear in the Astrophysical Journal Letters}
\received{8/20/2008}
\revised{9/9/2008}
\accepted{9/11/2008}

\shorttitle{Cometary Astropause of Mira} 
\shortauthors{T.\ Ueta}

\begin{document}
 
\title{Cometary Astropause of Mira Revealed in the Far-Infrared} 

\author{Toshiya Ueta}

\affil{Department of Physics and Astronomy,
University of Denver, 
Denver, CO 80208}

\email{tueta@du.edu}

\begin{abstract}
 Evolved mass-losing stars such as Mira enrich the interstellar medium
 (ISM) significantly by their dust-rich molecular wind. When these stars
 move fast enough relative to the ISM, the interaction between the wind
 and ISM generates the structure known as the astropause (a stellar 
 analog of the heliopause), which is a cometary stellar wind cavity 
 bounded by the contact discontinuity surface between the wind and
 ISM. Far-infrared observations of Mira spatially resolve the structure
 of its astropause for the first time, 
 distinguishing the contact surface between Mira's wind and the ISM and the 
 termination shock due to Mira's wind colliding with the ISM. The
 physical size of the astropause and the estimated speed of the
 termination shock suggest the age of the astropause to be about $40,000$
 yr, confirming a theoretical prediction of the shock
 re-establishment time after Mira has entered the Local Bubble.
\end{abstract}


\section{Introduction}

Mira is the archetype of asymptotic giant branch (AGB) stars, which
contribute significantly to the enrichment of the interstellar medium
(ISM) by a dust-rich molecular stellar wind (e.g.\ \citealt{sedlmayr94}). 
Hence, tracing mass-loss histories of these stars has a profound impact
on knowledge of the composition of the ISM and its evolution in host
galaxies.  
Mira is actually a wind-accreting binary system \citep{rc85}: a
mass-losing AGB star is the primary star (Mira A), dominating the
system.   
Mira's wind is, therefore, a cool molecular wind from the primary, in
which dust grains convert radiation pressure of the star into the
outward momentum of the wind \citep{kwok80,sss98}.  

Previous investigations in CO line emission revealed that Mira's wind
at $\sim 4$ km s$^{-1}$ formed a circumbinary molecular envelope of
$0\farcm3$ in radius, whose otherwise spherical structure
was punctuated by bipolar outflows at $\sim 8$ km s$^{-1}$
\citep{jos00,fong06}.   
Mira's space velocity is approximately $150$ km s$^{-1}$ nearly due
south as calculated via a spherical trigonometric transformation
\citep{js87} using its proper motion \citep{hip}, radial velocity
\citep{evans67} and the solar motion \citep{db98}. 

Hence, Mira's wind has been interacting with the ISM flow local to Mira
for quite some time.  
This interaction left a turbulent wake of approximately 2 degrees long, 
observed in emission in the ultraviolet \citep{m07} and 21 cm \ion{H}{1}
line \citep{m08}. 
The ultraviolet emission is thought to be due to H$_2$ excited
collisionally by hot electrons arising from the shocked ISM flow. 
The 21 cm line emission of atomic hydrogen is, therefore, expected from 
dissociation of H$_2$ into H.

This is the second case of the wind-ISM interaction around an AGB star
after R Hya, in which a bow shock resulting from the interaction was
seen in the far-infrared \citep{ueta06}.  
While the H$_2$/\ion{H}{1} wake of Mira has been attributed to the
wind-ISM interaction owing to Mira's large space velocity, distinct
shock layers of the interaction themselves have not been confirmed
observationally, except for the possible bow shock structure seen in
the ultraviolet \citep{m07}.
Thus, it has been unclear how the H$_2$/\ion{H}{1} turbulent wake seen 
at the large spatial scale is physically related to the CO wind seen at   
the small spatial scale, except that bright H$_2$ streams \citep{m07}
are correlated spatially with CO bipolar outflows \citep{jos00}.

In this letter, we present Mira's far-infrared image taken with the
{\sl Spitzer Space Telescope} ({\sl Spitzer}; \citealt{sst}), which
resolves spatially the structure of Mira's astropause (a stellar
analog of the heliopause; \citealt{nf82}) for the first time.
Resolved are 
(1) Mira's circumbinary dust envelope at the core of the astropause,
(2) the termination shock where the stellar wind is slowed down by the
ISM and 
(3) the astrosheath in which a turbulent wake assumes the characteristic
cometary shape beyond the termination shock.  
Thus, the {\sl Spitzer} $160\micron$ data, in conjunction with previous 
multi-wavelength results, provide the most comprehensive view of the
wind-ISM interaction for an AGB star to date. 

\section{Observations and Data Reduction}

Mira happened to be located at the northern edge of the {\sl XMM}-LSS
field in the {\sl Spitzer} Wide-Area Infrared Extragalactic Legacy
Survey (SWIRE; \citealt{swire}).
Far-infrared mapping of Mira and its vicinity was performed on 2004
August 1 and 2 with the Multiband Imaging Photometer for {\sl Spitzer}
(MIPS; \citealt{mips}).
Due to the relative position offset among the MIPS detectors in the
focal plane, Mira was scanned over only in the 24 and $160\micron$ bands 
(scans 08 and 09; AOR KEYs 5844992, 5845248, 11774976 and 11775232).
These scans were done at the medium rate along five scan legs of
3$^{\circ}$ long each with a cross-scan stepping of 148$\arcsec$.
Quick inspection of the archived data indicated that Mira was saturated
in both of the 24 and $160\mu$m bands: however, the $160\mu$m data
appeared salvageable.

Data reduction was done in the following steps with the Ge Reprocessing
Tools (GeRT; ver.\ 20060415 of S14 processing)\footnote{Available from
http://ssc.spitzer.caltech.edu/mips/gert/.} and the the Mosaicker 
software (ver.\ 20070615).\footnote{Available from
http://ssc.spitzer.caltech.edu/postbcd/.}   
First, we used GeRT to create our own basic calibrated data (BCD).
Bright sources can corrupt stimulator-response calibration if they
happen to be on the detector when the calibrating stimulator is
illuminated \citep{gordon}.
Thus, we excluded stimulator frames affected by bright Mira to optimize
the time-dependent detector responsivity calibration for our particular
data set.  
We also adjusted the number of reads in the data ramp to recover valid
data in otherwise saturated pixels.
Then, we applied a high-pass time median filter with a 50-detector-pixel
width on the custom-made BCDs to remove the ``streakings'' due to
residual slow-response variations of the detector, while masking out a
circular region around Mira of 30-detector-pixel radius to avoid
filtering out the extended structure around Mira. 
Finally, the custom-processed BCDs were mosaicked into a single map with
the Mosaicker.

The MIPS $160\micron$ band is known to suffer from a short-wavelength
light leak, which can result in a false point source offset from the
position of the true $160\micron$ source \citep{gordon}. 
Inspection of the mosaicked Mira map indicated that the emission core
was elongated by the presence of a secondary point source
offset from Mira (Figure \ref{map}a). 
Since Mira's $J$ magnitude is $-0.7$ (more than 6 magnitudes above the
limit over which the leak becomes detectable), the core elongation was
most likely due to the leak. 
Hence, we removed the light leak in the Mira map by scaling and
subtracting the image of a calibration K2III star, HD 163588, whose
$160\micron$ emission was known to be mostly the leak\footnote{MIPS Data
Handbook Ver.\ 3.3.1 (http://ssc.spitzer.caltech.edu/mips/dh/), p.76.}.
This yielded a single-peaked, PSF-like emission core. 
This method would inevitably subtract the photospheric
component of HD 163588 (estimated to be 56 mJy) from Mira. 
However, it was deemed negligible with respect to the flux of Mira at
$160\micron$ (52 Jy).  

\section{Results and Discussion}

\begin{figure}
 \plotone{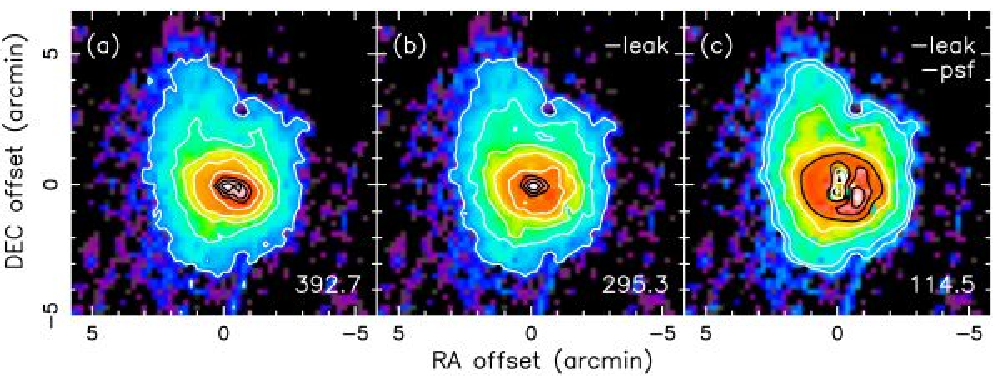}
 \caption{\label{map}%
 {\sl Spitzer}/MIPS far-infrared image of Mira's astropause at
 160\micron. 
 {\bf (a)} Calibration-optimized, saturation-corrected image in the
 12\arcsec pixel$^{-1}$ scale () centered at the position of Mira (north is up, east to the left)
 based on SWIRE/{\sl XMM}-LSS scans, which yielded 1 $\sigma$
 sensitivity of 0.8 MJy sr$^{-1}$.  
 The contours represent 80, 60, 40, 20, 10, 5, 2.5 and 0.5\% of the peak
 surface brightness, which is indicated at the bottom right corner in
 MJy sr$^{-1}$. 
 The lowest contour corresponds to emission at 3 $\sigma$ level ($= 2.5$
 MJy sr$^{-1}$). 
 The extent of the detected ($> 3 \sigma$) emission halo is 8\farcm5
 long and 6\farcm5 wide roughly at the PA of 15$^{\circ}$. 
 {\bf (b)}
 Same as (a), but for the short-wavelength light-leak removed image. 
 The slightly-elongated, singly-peaked core ($> 40\%$ of the peak) is
 surrounded by the similarly elongated plateau ($> 5\%$ of the peak) at
 the PA of 70$^{\circ}$. 
 {\bf (c)} Same as (b), but for the core-fitted-PSF subtracted
 image. 
 The lowest contour is $2.5\%$ of the peak. 
 The plateau is not entirely due to PSF and there is a ring of emission
 (at $\sim 40\%$ of the peak; $2\farcm4 \times 2\farcm0$ at the PA of 70$^{\circ}$).}    
\end{figure}

Figure \ref{map} displays versions of the {\sl Spitzer}/MIPS image of
Mira at $160\micron$ in log-scaled false-color.
These maps were made at the $12\farcs0$ pix$^{-1}$ scale ($11\farcm7
\times 13\farcm3$ field of view) with the mean sky coverage of $3.5 \pm
1.9$ pixel$^{-1}$.  
The effect of saturation was not removed entirely by adjusting reads in
the data ramp, and the $12\farcs0$ pix$^{-1}$ scale turned out to be the
minimum pixel scale for which all pixels register valid data.
The one $\sigma$ sensitivity and the average large-spatial-scale sky
emission (the component removed during the median filter) were found to 
be 0.8 and $8.8 \pm 1.0$ MJy sr$^{-1}$, respectively.
The derived sky emission value was consistent with the estimated value of 6.5 
MJy sr $^{-1}$, which is obtained with the {\sl Spitzer} Planning
Observations Tool (SPOT)
software.\footnote{http://ssc.spitzer.caltech.edu/documents/background/} 
Contours are 80, 60, 40, 20, 10, 5, 2.5 and 0.5$\%$ of the emission
peak, with the lowest contour corresponding to 3 $\sigma$ ($= 2.5$ MJy
sr$^{-1}$).  

The light-leak-corrected image is presented in Figure \ref{map}b, in
which the emission core is evidently single-peaked. 
The leak-corrected emission profile of Mira can be dissected into three
distinct emission regions: the emission core, plateau and halo. 
The core is slightly extended with the full-width at half-maximum
(FWHM) of $1\farcm0 \times 0\farcm9$ at the position angle (PA; measured
East from North) of
70$^{\circ}$. 
Since nearby point sources have the average FWHM of $0\farcm7$, the
extension of the core appears genuine. 
The plateau is an elliptical region ($3\farcm4 \times 2\farcm8$ at the 
PA of 70$^{\circ}$) of relatively flat profile encircling the core (at 5\% 
to 30\% of the peak emission). 
The halo is a fainter nebulosity surrounding the plateau. 
If we define the periphery of the halo at 3 $\sigma$ (the lowest contour
in Figure \ref{map}), then its extent is $8\farcm5 \times 6\farcm5$ at 
the PA of 15$^{\circ}$. 
The scan-map direction is the PA of 160$^{\circ}$, along which some
residual ``streakings'' are still recognizable below 3 $\sigma$. 
Hence, the elongation of the $160\micron$ emission around Mira is not
the artifact of scan mapping. 
Mira has been imaged in the far-IR most recently with the {\sl AKARI 
Infrared Astronomy Satellite} \citep{murakami07}.
{\sl AKARI}'s Mira maps are consistent with the {\sl Spitzer} map with a
single-peaked emission core with extension at the PA of 15$^{\circ}$
(H.\ Izumiura et al.\ 2008, private communication). 
 
The point-spread-function (PSF) in the MIPS $160\micron$ band has the
first Airy pattern out to $\sim 7\arcmin$ (at $12\arcsec$ pixel$^{-1}$)
at $\sim 4\%$ of the peak surface brightness. 
A synthesized PSF\footnote{%
Available at
http://ssc.spitzer.caltech.edu/archanaly/contributed/stinytim/.} was
convolved with an elliptical Gaussian function to conform the observed
shape of the core and subtracted from the data (Figure \ref{map}c). 
We immediately see that there is still substantial emission ($\sim 50$ MJy
sr$^{-1} = \sim 20 \sigma$) in the plateau, suggesting that it is caused
by an elliptical ring of $2\farcm4 \times 2\farcm0$ at the PA of
70$^{\circ}$. 
The core is oriented in the same way as the circumbinary CO envelope
\citep{jos00}, while the CO envelope is smaller than the
core \citep{fong06}. 
At the distance of 107 pc to Mira \citep{knapp03}, the physical size 
of the CO envelope is $\sim 3 \times 10^{16}$ cm in radius. 
This agrees well with the expected radius of Mira's molecular envelope
resulted from mass loss at the rate of $\sim 2 \times 10^7$ M$_{\odot}$
yr$^{-1}$ \citep{jos00,rs01,fong06}, including the effect of 
photodissociation by the interstellar radiation field \citep{mamon88}.  
Therefore, the core represents Mira's circumbinary dust envelope whose
physical size is $\sim 4.5 \times 10^{16}$ cm in radius. 

\begin{figure}
 \begin{center}
  \epsscale{0.85}
  \plotone{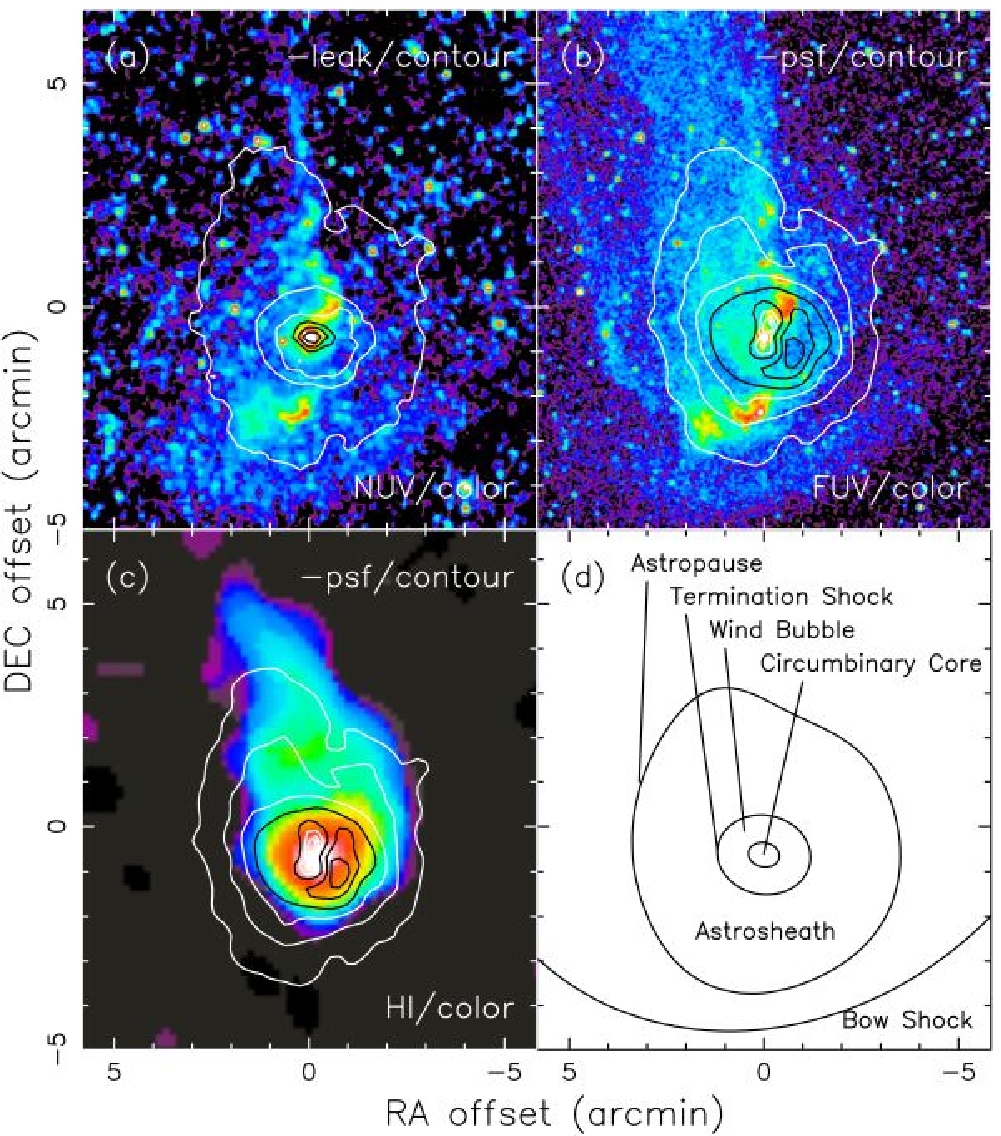}
 \end{center}
 \caption{\label{over}%
 H$_2$/\ion{H}{1} imaging data overlaid with the $160\micron$ contours of
 Mira's astropause, and a schematic of its structure emerging from the
 present study. 
 {\bf (a)} Near-ultraviolet color image \citep{m07} overlaid with the
 leak-corrected $160\micron$ contours, showing the spatial relationship
 between emission  in these two bands. 
 The contours represent 80, 60, 40, 20, 10, and 0.5\% (3 $\sigma$) of   
 the peak surface brightness. 
 {\bf (b)} Same as (a), but far-ultraviolet color image \citep{m07}
 overlaid with the leak-corrected, core-subtracted $160\micron$ contours.  
 {\bf (c)} Same as (a), but 21 cm \ion{H}{1} color image \citep{m08}
 overlaid with the leak-corrected, core-subtracted $160\micron$ contours. 
 {\bf (d)} Schematic of the structure of Mira's astropause revealed by
 the $160\micron$ image and its spatial relationship with respect to
 H$_2$/\ion{H}{1} distribution. 
 The circumbinary dust envelope, the termination shock and the astrosheath
 are spatially distinguished in the far-infrared for the first time.}    
\end{figure}

Since the Mira system is moving at $\sim 150$ km s$^{-1}$
the resulting wind-ISM interaction has given rise to the ultraviolet bow 
shock structure \citep{m07}, despite Mira's slow wind velocity of 4 km
s$^{-1}$ \citep{jos00,fong06}.  
Figures \ref{over}a and \ref{over}b clearly show that the downstream
side of the ultraviolet bow shock coincides with the southern edge of
the far-infrared halo. 
Moreover, the cometary shape of the halo is the most consistent with the
bow shock interpretation, in which the far-infrared nebulosity is shaped
by a turbulent wake flowing almost due north. 
According to the theory of the (solar)wind-ISM interaction, the ISM flow
approaches supersonically toward Mira, becomes subsonic past the bow
shock and streams around the astropause, while Mira's wind expands
freely to form a bubble bounded by the termination shock, beyond which
the wind becomes compressed and turbulent and flows downstream in the
astrosheath \citep{zank99}.  
Assuming that the $160\micron$ emission probes Mira's dusty wind
emanating from the circumbinary envelope, we interpret that the
leak-corrected, core-subtracted image distinguishes spatially all
components of the astropause (Figure \ref{map}c) with the halo tracing
the astrosheath and the emission ring delineating the termination shock
(see, also, Figure \ref{over}d).  

The prominent ultraviolet streams (Figures \ref{over}a and \ref{over}b)
are outflows of H$_{2}$ excited collisionally by hot electrons in the
bow-shock-excited ISM flow \citep{m07}. 
These H$_{2}$ flows, therefore, must have penetrated the termination
shock and been directed downstream while getting excited in the
astrosheath. 
In the astrosheath, H$_2$ can be dissociated by either the interstellar
radiation field \citep{mamon88} or collisional excitation
\citep{m07,m08}. 
The observed \ion{H}{1} emission is mostly concentrated in the vicinity
of the termination shock, with the peak FWHM of $1\farcm8$ \citep{m08},
which is compatible with the size of the molecular radius given the
resolution at 21 cm (Figure \ref{over}c). 
Similar to H$_2$, atomic H is directed toward downstream beyond the
termination shock, lending strong support for the multi-phase
characteristics of the flow in the astrosheath. 

The stagnation point of the wind-ISM interaction is therefore at the
apex of the halo (astropause), which is $3.6 \times 10^{17}$ cm from
Mira. 
Requiring ram pressure balance at the stagnation point, this implies
that the ISM density local to Mira is $n_{\rm ISM} = 0.02$ cm$^{-3}$. 
This value is consistent with the result of a two-wind model
\citep{wareing07} that accounts for Mira's entry into the Local Bubble,
the region of warm ($\sim 10^{6}$ K), tenuous ($\sim 0.01$ cm$^{-3}$) ISM in
the solar neighborhood \citep{l01}. 
Given $N_{\rm H I} \le 1.3 \times 10^{19}$ cm$^{-2}$ and 25\%
abundance of neutral species \citep{m08}, we can estimate the optical
depth at $160\micron$ to be $8 \times 10^{-6}$ assuming a power-index
scaling law with the index of 1.2 (i.e. silicate dust). 
Under the assumption that thermal dust emission dominates at
$160\micron$, the emission map and the optical depth yield the
temperature of dust; $\sim 500 - 1100$ K in the circumbinary dust
envelope, $\sim 200$ K in the termination shock and $\sim 30 - 200$ K in
the astrosheath. 

Based on these estimates, the pre-shock temperature of Mira's wind is
thought to be $\sim 50$ K (assuming gas-dust thermalization), implying
Mach number of 7.5. 
With the generalized jump relations (e.g.\ \citealt{p81}), the velocity
of the termination shock is $-2$ km s$^{-1}$ in Mira's frame and the
ratio of post- to pre-shock densities 3.8. 
These values yield the post-shock gas temperature of $\sim 900$ K. 
Thus, in the shock-excited astrosheath, dust is still present (the gas
temperature is below the condensation temperature of silicates $\approx
1200$ K; \citealt{s77}), but probably de-thermalized from gas. 
Spectroscopic follow-ups are necessary to investigate the evolution of
the dust-gas characteristics upon passage of the termination shock. 

At 2 km s$^{-1}$, the termination shock should have taken $\sim 40,000$ 
yr to advance to the present position ($1 \times 10^{17}$ cm) from the
astropause ($3.6 \times 10^{17}$ cm). 
The total flux of the astrosheath (including the termination shock) is
37 Jy, and this translates to the total mass of $7 \times 10^{-3}$
M$_{\odot}$ using the dust emissivity of generic interstellar 
dust \citep{dl07}.  
With Mira's mass loss rate ($\sim 2 \times 10^{-7}$ M$_{\odot}$
yr$^{-1}$; 
\citealt{jos00,fong06,rs01})\footnote{Various authors quote a range of
mass-loss rates for Mira, which is a variable star. In the present
paper, however, we are concerned with the long-term effects of Mira's
mass loss.  Therefore, we adopt the average rate of mass loss in our
discussion.},  
it should have taken $\sim 40,000$ yr for the astrosheath to form. 
These estimates for the age of the termination shock is consistent with
the time needed for the termination shock to re-establish itself after
entering the Local Bubble ($\sim 40,000$ yr) as estimated numerically by
\citet{wareing07}, who attributed a kink in Mira's ultraviolet tail
\citep{m07} to Mira's entry into the Local Bubble.

The presence of the bow shock indicates that the outer shock is
supersonic. 
In the warm ($\sim 10^6$ K), tenuous ($\sim 0.01$ cm$^{-3}$) Local
Bubble \citep{l01}, however, the shock velocity of $\sim 150$ km
s$^{-1}$ is barely supersonic. 
This suggests that the pre- and post-shock parameters across the bow
shock are nearly continuous. 
This is consistent with the fact that the ultraviolet emission profile
does not indicate any density enhancement (i.e.\ material pile-up) at
the southern edge of the bow. 
Hence, ultraviolet emission in front of the upstream face of the
astropause is due to excitation of the post-shock ISM gas by a shear
flow. 
While Mira's wind and the ISM gas flow are in general separated by the
astropause, these flows are inherently turbulent. Thus, these flows are
mixed in the downstream wake and H$_2$ is collisionally excited,
resulting in the observed far-ultraviolet emission. 
Excited H$_2$ can be dissociated, leading to the observed \ion{H}{1}
emission. 
Dust grains help H$_2$ to replenish in the wake, and this cool,
turbulent three-phase gas flow results in emission of H$_2$ and
\ion{H}{1} continuing at least for 1.5$^{\circ}$ long. 

Hence, there is an emerging picture of Mira's astropause (Figure
\ref{over}d). 
The extent of the far-infrared emission outlines the astropause, which
is the contact surface between Mira's molecular wind and the ISM flow. 
The bulk of the halo emission is due to dusty molecular wind material in
the astrosheath, shock-excited by the termination shock delineated by
the emission ring.
The termination shock is the discontinuous interface between the pre-
and post-shock regions of Mira's wind, inside which Mira's wind 
expands undisturbed coming off the circumbinary envelope hosting bipolar
outflows.

Since the post-shock regions of both Mira's wind and the ISM flow have a
significant physical size, the shock must be non-radiative (i.e., it has
not been cooled down and compressed into an unresolvable size).
According to \citet{cloudy}, in a collisionally excited solar wind
abundance gas heating dominates cooling below 10$^4$ K (their Fig.\ 6).
The estimated post-shock gas temperature of Mira's wind is $\sim 900$
K. 
Thus, the post-shock gas can indeed maintain its temperature, i.e.,
its physical size.
Because previous numerical models (\citealt{wareing07,raga08}) were
concerned mainly with the structure of the turbulent wake of the
wind-ISM interaction, follow-up multi-phase hydrodynamical modeling at
low temperatures to reproduce the structure of the astropause appears
extremely worthwhile. 

The total flux of the astropause is 52 Jy, which implies the total mass
of $1 \times 10^{-2}$ M$_{\odot}$ using the dust emissivity of generic
interstellar dust \citep{dl07}. 
Since the total atomic mass in the astropause is $1.3 \times 10^{-3}$
M$_{\odot}$ \citep{m08}, the total molecular mass is estimate to be 
$8.7 \times 10^{-3}$ M$_{\odot}$ to be roughly 1 to 9 atomic-to-molecular
mass ratio. 
The observed far-ultraviolet luminosity implies the H$_{2}$ dissociation
rate of $\sim 2.5 \times 10^{42}$ s$^{-1}$, if collisional
excitation accounts for all far-ultraviolet emission \citep{m07}. 
This rate, however, would have yielded a factor of four more atomic H
than observed \citep{m08}. 
Dust grains can potentially provide H$_{2}$ formation sites to reduce
excess atomic H. 
Using parameters of gas in the astrosheath ($n_{\rm H} \approx 7$ and
$T_{\rm gas} \approx 900$ K) and silicate grains ($0.1\micron$ radius,
3.7 g cc$^{-1}$, 1\% mass ratio), the H$_2$ formation rate is
approximately $4 \times 10^{40}$ s$^{-1}$, if one assumes that 
an encounter with two atomic H by a dust grain always results in
production of H$_2$. 
This suggests that H$_2$ far-ultraviolet emission is not entirely due to
collision of H$_2$ with hot electrons and by some other means as already
implied \citep{m07,m08} and/or there exist some mechanisms to bolster
production of H$_2$ from atomic H.  

Clearly, Mira's astropause presents a unique laboratory for
investigations of a multi-phase hydrodynamical flow, turbulence and
dust-gas processing, which has a profound impact on the chemical makeup 
of the ISM. 
Because recent findings suggest that wind-ISM interaction around AGB
stars may be common \citep{ueta06,ueta07}, spatially resolved
spectroscopy of multi-phase dusty, low-temperature gas flow will be
extremely productive in the coming era of Herschel, SOFIA and ALMA
(e.g.\ \citealt{herschel,sofia,alma}, respectively).

\acknowledgements
Ueta thanks T.\ Le Bertre and L.\ D.\ Matthews for making their
\ion{H}{1} data available and R.\ E.\ Stencel for careful reading of the
manuscript. 
This work is based on archival data obtained with the Spitzer Space
Telescope, which is operated by the Jet Propulsion Laboratory,
California Institute of Technology under a contract with NASA. 
Support for this work was provided partially by an award issued to the 
University of Denver by JPL/Caltech.

\end{document}